\documentclass[12pt]{article}
\advance\hoffset by -7mm  
\makeatletter
\@addtoreset{equation}{section}
\makeatother

\renewcommand{\title}[1]{\null\vspace{25mm}

\noindent{\Large{\bf #1}}\vspace{10mm}

\noindent {\large By }}
\newcommand{\authors}[1]{\noindent{\large #1}\vspace{3mm}

}
\newcommand{\address}[1]{\noindent #1\vspace{5mm}

}
\renewcommand{\abstract}[1]{\vspace{19mm}

\noindent{\small{\em Abstract.} #1}\vspace{2mm}

} 
\usepackage{amstex,righttag}
\begin{document}
\title{Inflation with a Complex Scalar Field}
\authors{David Scialom}
\address{Institute of Theoretical Physics, University of Z\"urich,\\
Winterthurerstrasse 190, CH-8057 Z\"urich, Switzerland}
\abstract{\noindent We discuss
the coupled Ein\-stein-Klein-Gor\-don equations for
a complex scalar field with and 
without a quartic self-interaction in a zero curvature
Friedman-Lema\^{\i}\-tre Universe. The complex scalar
field, as well as the metric, is decomposed in a homogeneous, 
isotropic part (the background) and in first order gauge invariant scalar 
perturbation terms.
The background equations can be written as a set of 
four coupled 
first order non-linear differential equations.
These equations are analyzed using modern theory of dynamical
system.
It is shown that,
in all singular points where inflation occurs,
the phase of the complex scalar field is asymptotically constant. 
The analysis of the first order equations is done for the
inflationary phase.
For the short wavelength regime the first order perturbation term of
the complex scalar field is smeared out and the Bardeen potential
oscillates around a nearly constant mean value. Whereas for the long
wavelength regime the first order perturbed quantities increase.}
%\vskip -2cm
%\hbox{ }\hfill
\section{Basic equations}
We will consider the linear scalar mode perturbations since 
they are the only 
ones which contribute to the energy density fluctuations \cite{kod}. We will, 
as usual, expand the scalar perturbations 
in terms of a complete set of harmonic functions 
$Y_k$, which are the eigenfunctions with eigenvalue $-k^2$ of the
Laplace-Beltrami operator 
$\mathbf{\Delta}$ defined on constant time slices $\Sigma_{\eta}$.   
In the following, we omit for simplicity to write the 
subscript $k$ on $Y$.\\
\indent
Taking the zero curvature
Fried\-mann\--Le\-ma\^{\i}\-tre metric---including the first order
scalar mode perturbations---given in Ref.~\cite{kod} and using the
following action
\begin{equation}
  	S=\int \sqrt{-g}d^4x \left[-\frac{R}{16\pi G}+
	\frac{1}{2}g^{\mu\nu}\left(e_{\mu}\Phi 
  	e_{\nu}\Phi^{\ast}+e_{\nu}\Phi e_{\mu}\Phi^{\ast}\right)+
  	m^2\Phi\Phi^{\ast}+\lambda\left(\Phi\Phi^{\ast}\right)^2
%	V\left(\Phi,\Phi^{\ast}\right)
	\right],
 	\label{smat}
\end{equation}
the equations of motion of the background as well as the first order
perturbations equations can be derived. To obtain the latter, the scalar
field has to be expanded in a background part, $\phi$ and in a first
order term, $\delta\phi$.
%In eq.(\ref{smat}) $\zeta$ is the volume form  and
%$V=m^2\Phi\Phi^{\ast}+\lambda\left(\Phi\Phi^{\ast}\right)^2$.
We get for the background
\begin{gather}
H^{2}=\frac{8\pi G}{3}(\dot{\phi}
\dot{\phi}^{\ast}+m^{2}\phi \phi^{\ast}+\lambda
(\phi \phi^{\ast})^{2})~,   \label{e00}\\
-2\dot{H}-3H^{2}=8\pi G
( \dot{\phi}\dot{\phi}^{\ast}-m^{2}\phi \phi^{\ast}-
\lambda (\phi \phi^{\ast})^{2})~.
\label{eij}\\
\ddot{\phi}+3H\dot{\phi}+m^{2}\phi+2\lambda (\phi
\phi^{\ast})\phi=0~,  \label{kg}
\end{gather}
where $H=\dot{a}/a$ and dot means derivative with respect to the 
real time, $t$.\\
\indent
Expressing the first order perturbation equations directly with
respect to gauge invariant quantities \cite{bar} we obtain
\begin{gather}
	6h\psi^{'}+\left(2k^2+6h^2\right)\psi = 8\pi G \biggl(\phi^{\ast '}
	\delta\varphi^{'}+
	\phi^{'}\delta\varphi^{\ast '}
	+2\phi^{'}\phi^{\ast '}\psi
	+a^2\frac{\partial V}{\partial\phi}\delta\varphi+
	a^2\frac{\partial V}{\partial\phi^{\ast}}\delta\varphi^{\ast}\biggr),
	\label{de00}\\
	\psi^{'}+h\psi = -4\pi G\left(\phi^{\ast '}\delta\varphi+
	\phi^{'}\delta\varphi^{\ast}\right),
	\label{de0i}\\
	\psi^{''}+6h\psi^{'}+\left(k^2+16\pi Ga^2 V\right)\psi =
        8\pi Ga^2\left(\frac{\partial V}{\partial\phi}\delta\varphi+
        \frac{\partial V}{\partial\phi^{\ast}}\delta\varphi^{\ast}\right),
 	\label{deij}\\
	\delta\varphi^{''}+2h\delta\varphi^{'}+k^2\delta\varphi+
	4\phi^{'}\psi^{'}-
	2a^2\frac{\partial V}{\partial\phi^{\ast}}\psi+
	a^2\frac{\partial^2 V}{\partial\phi^{\ast 2}}\delta\varphi^{\ast}+
	a^2\frac{\partial^2 V}{\partial\phi^{\ast}\partial\phi}\delta\varphi=0,
	\label{dkg}
\end{gather}
where $\psi$ is the Bardeen potential, $h=a^{'}/a$, $\delta\varphi$
is the gauge
invariant quantity corresponding to $\delta\phi$ and prime means
derivation with respect to the conformal time, $\eta$.\\
\indent
Since the matter action is U(1)-globally invariant we get the
conservation with respect to the conformal time of the bosonic charge
\begin{equation}
\Xi=\frac{i\, a^2}{2}\left(\phi^{\ast '}\phi-
\phi^{'}\phi^{\ast}\right)\, .
\label{Xi}
\end{equation}
It should be noticed that the first order terms are entirely determined by
eqs.(\ref{de00})-(\ref{de0i}), eq.(\ref{dkg}) and their complex conjugate. 
The background solution is
established by eq.(\ref{e00}), eq.(\ref{kg}) and their complex conjugate.
\vskip -2cm
\hbox{ }\hfill
\section{The background}
\hbox{ }\hfill
\vskip -2cm
We will only consider here the case where $m\neq 0$. For a complete
discussion of the background see Ref.\cite{sci}.
The only singular point not lying at infinity of the phase
space is the coordinate origin. This singular point is an
asymptotically stable winding point. It correspond to the oscillatory
phase of the complex scalar field.\\
\indent
In order to find the singular points lying at infinity we have to
perform a transformation which maps them on the boundary of a unit
three-sphere. We extend the phase space to the boundary to analyze
their behavior.
It turns out that, whatever the value of $\lambda$ is,
there is a line of singular points
lying at infinity which meet the criteria of inflation. For $\lambda=0$,
the asymtotical behavior of the scalar field and the Hubble parameter near
line of singular points is given by
\begin{equation}
\varphi= \frac{-M_{p}mt}{\sqrt{3}}~e^{i\vartheta_{30}}~, 
\qquad
H= \frac{-m^{2}t}{3}\label{eql03}~,
\end{equation}
for $t\rightarrow\mbox{}-\infty$. Setting $\vartheta_{30}=0$, we
recover the result found in Ref.\cite{bel}.
Similarly, for the case $\lambda\neq 0$,
we have
\begin{equation}
\varphi= \varphi_0 e^{i\vartheta_{30}}
\exp\left(-2M_p\sqrt{\frac{\lambda}{3}}~t\right)~, \label{LL1}
\end{equation}
where $t\rightarrow -\infty$ and $\varphi_0$ is a negative integration 
constant. Using eq.(\ref{e00}), one gets the asymptotic behavior of the
Hubble parameter. 
We see that the above given asymptotic solutions
correspond to outgoing separatrices in phase space.
The fact that along these separatrices the
phase of $\varphi$ remains constant is important and shows that
inflation is essentially driven by one component of the field.
Notice that this conclusion is also valid for the massless case.
\vskip -2cm
\hbox{ }\hfill
\section{Perturbation during inflation}
\hbox{ }\hfill
\vskip -2cm
We will consider the
long wavelength and the short wavelength limits separately. To solve the
Einstein equations, we first define the complex valued function
$U(\eta)$ as the solution of the following system of differential equations
\begin{gather}
U^{'}+hU=-4\pi G\phi^{\ast '}\delta\varphi ,\label{a1}\\
U^{''}+2\left(h-\frac{\phi^{\ast ''}}{\phi^{\ast '}}\right)U^{'}
+\left(k^2+2h^{'}-2h\frac{\phi^{\ast ''}}{\phi^{\ast '}}\right)U=0.
\label{a2}
\end{gather}
To simplify the notation
we omit to write the explicit k dependence on $U$.
As long as \hbox{$|\phi^{\ast '}|
\lower 4pt \vbox{\vskip 3pt\hbox{$>$}\vskip -8.82pt \hbox{$\sim$}}
\left|\delta\varphi\right|$}, one easily sees that the sum $(U+U^{\ast})$ 
fulfills the two Einstein eqs.(\ref{de00})-(\ref{de0i}) and thus can
be identified with $\psi$.
Setting $U=\frac{\phi^{\ast '}u}{a}$,
$g=\frac{h}{a\phi^{\ast '}}$ and using the background equations,
eq.(\ref{a2}) can be rewritten as
\begin{equation}
u^{''}+k^2u+\left[\frac{1}{h}\left(
\frac{\phi^{\ast ''}}{\phi^{\ast '}}-
\frac{\phi^{''}}{\phi^{'}}\right)
\left(h^2-h^{'}\right)-\frac{g^{''}}{g}\right]u=0.
\label{u22}
\end{equation}
By solving this last equation
we obtain $\psi$ and $\delta\varphi$.
In order to solve some problems of the standard cosmological model 
(e.g. the flatness problem ...) a sufficiently long inflationary stage is
needed. Hence, inflation starts with $\vartheta$ close
to a constant.
Writing
$\phi=\left|\phi\right|e^{i\vartheta}$, we get from eq.(\ref{Xi})
\begin{equation}
a^2{\left|\phi\right|}^2\vartheta^{'}=\Xi, \label{Xit}
\end{equation}
where $\Xi$ is the constant bosonic charge.
On the separatrix, where inflation
occurs, we have, whatever the values of $m$ and $\lambda$ are,
that $a^2{\left|\phi\right|}^2$ is
growing exponentially. From eq.(\ref{Xit}) we see that,
immediately after the beginning of inflation,
$\vartheta^{'}$ can be taken to be
zero within first order approximation. Thus,
$\vartheta$ will be 
constant as long as inflation lasts. As a consequence eq.(\ref{u22}),
reduces to
\begin{equation}
u^{''}+k^2u-\frac{g^{''}}{g}u=0.
\label{u22i}
\end{equation}
Since during inflation, $a^2H\gg g{''}/g$, we get immediately for wavelength
perturbations smaller than the Hubble radius that the dominant terms
of $\psi$ and $\delta\varphi$ are given by \cite{jet,sci2}
\begin{gather}
\psi =2 Re\left[\dot{\phi}\left(\gamma_1\cos\left(k\int\frac{dt}{a}\right)+
\gamma_2\sin \left(k\int\frac{dt}{a}\right)\right)\right],
\label{psisw}\\
\delta\varphi \simeq -\frac{k}{4\pi G\,a}
\left[\gamma_2^\ast\cos (k\eta)-\gamma_1^\ast\sin
(k\eta)\right],\label{phisw}
\end{gather}
where $Re$ denotes the real part and $\gamma_1, \gamma_2$ are
complex integration constants.
The slow-rolling approximation,
required for having a sufficiently long
inflationary stage, leads to a slowly variation of the mean value of
$\psi$.
The behavior of $\delta\varphi$ is governed by the $1/a$
factor, which decreases rapidly. It follows, as expected, that the
short wavelength fluctuations of the scalar field are smeared out.\\
\indent
For  $k\ll g^{''}/g$---the long wavelength perturbations--- 
eq.(\ref{u22i}) can also be solved. Hence, we obtain for the dominant terms  
\begin{equation}
\psi
= 2 Re\left(\tilde{k}\right)\left[1-\frac{H}{a}\int^ta\,dt\right]
\simeq -2 Re\left(\tilde{k}\right)\frac{\dot{H}}{H^2},
\hskip 0.5cm
\delta\varphi \simeq -\frac{k_2\dot{\phi}^\ast}{4\pi G\,H}.\label{plw}
\end{equation}
with $k_2$ and $\tilde{k}=k_2e^{-2i\vartheta}/8\pi G$ being complex
integration constants.
We consider an initial perturbation with wavelength inside the Hubble radius,
which will be outside it at the end of
inflation. There are wavelengths that fulfill
these conditions.
At the end of inflation, the evolution of the gauge invariant metric
potential is given by eq.(\ref{plw}). Later, when the universe is
dominated by relativistic particles, the scale factor scales as
$a\propto t^\mu$. Hence, the Hubble radius increases more rapidly than
the fixed comoving wavelength. The metric perturbation can re-enter
inside the Hubble radius and induce fluctuations on the ordinary
matter. At Hubble radius crossing, using eq.(\ref{plw}), the metric
perturbation is given by
\begin{equation}
\psi=2 Re\left(\tilde{k}\right)\frac{1}{\mu+1}.
\end{equation}
During the inflationary phase the behavior of
the perturbations is similar to the one of the real scalar field.
This is not surprising,
since along the separatrices the phase of the
complex scalar field remains constant and thus inflation is
essentially driven by one component of the field.
\vskip -2cm
\hbox{ }\hfill

\end{document}